\documentclass{jpsj2}
\newcommand{\Vect}[1]{{\textnormal{\mathversion{bold}$#1$}}}
\title{Exact Ground-State Energies of the Random-Field Ising 
Chain and Ladder}
\author{Toshiyuki Hamasaki and Hidetoshi Nishimori}
\inst{Department of Physics, Tokyo Institute of Technology,
Oh-okayama, Meguro-ku, Tokyo 152-8551}

\recdate{\today}

\abst{
 We derive the exact ground-state energy of the one-dimensional Ising
 model in random fields taking values $h$, 0 and $-h$ with general
 probabilities.
 The random-field Ising model on a ladder is also analyzed by showing
 its equivalence to the random-field Ising chain with field values 
 $\pm h$ and 0 for $h<J$.
 The zero-temperature transger matrix is used to obtain the results.
 }

\kword{
transfer matrix, random-field Ising model, exact solution, 
ground-state energy
}

\begin{document}
\maketitle

 \section{Introduction} 
 The one-dimensional random-field Ising model is one of the simplest
 examples of systems with quenched disorder.
 In the long history of its intensive studies, a number of results have
 been published which give exact solutions in the ground state.
 Derrida {\it et al}\cite{DVP78} first derived the exact expression of
 the ground-state energy by a recursive calculation of the
 zero-temperature transfer matrix.
 They also obtained the exact ground-state energies of the
 one-dimensional random bond Ising model in a uniform field, the random
 bond Ising ladder and the $3\times 3$ square lattice system.
 Farhi and Gutmann\cite{FG93} obtained the exact spin-spin correlation
 function of the one-dimensional random-field Ising model.
 Their approach is based on the ground-state spin configuration
 associated with the general rule that dictates which regions of the
 random field configurations necessarily determine the spin
 configuration.
 This method is related to another method which calculates exactly the
 energy and the entropy of the random-bond Ising model at zero
 temperature\cite{V78}.
 The application of the method proposed by Derrida {\it et al} to
 error-correcting codes was performed by Dress {\it et al}\cite{DAL95}.
 They investigated a simple error-correcting code through the
 zero-temperature properties of a related spin model.
 Further generalization of the method proposed by Dress {\it et al} was
 carried out by Kadowaki {\it et al}\cite{KNN96}.
 They calculated exactly the ground-state energies of the $\pm J$ random
 bond model and the site-random model on strips of various widths.
 Studies of the effects of continuous random fields in the ground state
 related to hysteresis are also found in
 Refs. [\citen{S96}][\citen{S00}][\citen{DDMZ02}].

 In the present paper, we derive the exact ground-state energy of the
 one-dimensional Ising model in random fields consisting three possible
 values $\pm h$ and 0 using a generalization of the methods in
 Refs. [\citen{DAL95}] and [\citen{KNN96}].
 The solution is piecewise linear as a function of the
 random-field strength, a property shared with the case of random fields
 $\pm h$.
 We also find that the one-dimensional Ising model in random fields
 $\pm h$ and 0 is equivalent to the Ising spin ladder in random fields
 $\pm h$ as long as $h$ is smaller than the exchange interaction $J$.
 This observation leads to the exact expression of the ground-state
 energy of the ladder under random field, which represents the first
 exact solution for the random-field Ising ladder.

 The present paper is composed of seven sections.
 In \S 2, we introduce our model and explain the formulation of our
 method based on the zero-temperature transfer matrix. 
 An important feature of the ground state of the one-dimensional Ising
 model in random fields is explained in \S 3.
 The explicit evaluation of the probability distribution of a relevant
 quantity and the exact ground-state energy are shown in \S 4 and \S 5.
 A useful property for diagonalization of the transition matrix is also
 given.
 In \S 6, we consider the Ising ladder in $\pm h$ random fields and
 show that this model is equivalent to the one-dimensional Ising model
 with three possible random fields $\pm h$ and 0 if $h<J$.
 The exact ground-state energy of the ladder model is derived using the
 result for the one-dimensional case.
 Finally in \S 7, results are discussed.

 \section{Ising spin chain in the general random field}

  \subsection{Model}
  Our problem is defined by the Hamiltonian
  \begin{eqnarray}
   H = -J\sum_{n=1}^{N}S_{n-1}S_{n} -\sum_{n=1}^{N}h_{n}S_{n},
    \label{eqn:RFIM}
  \end{eqnarray}
  where $J$ is the ferromagnetic coupling constant, $S_{n}(=\pm 1)$ is
  the Ising spin variable and $N+1$ is the number of spins.
  We use an open boundary condition.
  The random field at site $n$ consists of three possible values with
  probabilities
  \begin{eqnarray}
   P(h_{n}) =
    \left\{
     \begin{array}{ccl}
      p& \textrm{for}& h_{n}=+h \\
      q& \textrm{for}& h_{n}=0 \\
      r& \textrm{for}& h_{n}=-h \\
     \end{array}
	   \right.,
    \hspace*{1cm}
    p +q +r =1.\label{eqn:three}
  \end{eqnarray}
  Almost all of the previous studies concentrated on the case $q=0$.
  The partition function of this model with spin $S_{N}$ on the right
  edge fixed is
  \begin{eqnarray}
   Z_{N}(S_{N}) =
    \sum_{S_{0}} \sum_{S_{1}} \cdots \sum_{S_{N-1}}
    \prod_{n=1}^{N}\exp\{\beta(JS_{n-1}S_{n} +h_{n}S_{n})\},
    \label{eqn:ZN}
  \end{eqnarray}
  where $\beta$ is the inverse temperature.
  We can construct the partition function $Z_{N+1}(S_{N+1})$ recursively
  as 
  \begin{eqnarray}
   Z_{N+1}(S_{N+1}) =
    \sum_{S_{N}}Z_{N}(S_{N})\exp(\beta(JS_{N}S_{N+1} +h_{N+1}S_{N+1})).
    \label{eqn:PF}
  \end{eqnarray}
  To simplify this expression, we rewrite (\ref{eqn:PF}) by using a
  $2\times 2$  matrix.
  Let us introduce the notation $Z_{N}^{\pm} \equiv Z_{N}(S_{N}=\pm 1)$.
  The recursion relation (\ref{eqn:PF}) can then be written as
  \begin{eqnarray}
   \left[
    \begin{array}{c}
     Z_{N+1}^{+}\\
     Z_{N+1}^{-}\\
    \end{array}
   \right]
    =
    T_{N+1}
   \left[
    \begin{array}{c}
     Z_{N}^{+}\\
     Z_{N}^{-}\\
    \end{array}
   \right],
   \label{eqn:TM}
  \end{eqnarray}
  where $T_N$ is the transfer matrix
  \begin{eqnarray}
   T_{N} =
    \left[
     \begin{array}{cc}
      z^{J+h_{N}} & z^{-J+h_{N}}\\
      z^{-J-h_{N}} & z^{J-h_{N}}\\
     \end{array}
    \right], \label{eqn:matrix}
  \end{eqnarray}
  with $z = {\rm e}^{\beta}$.
  
  \subsection{Recursion relations in the zero-temperature limit}
  In the limit of very low temperatures, the partition function
  $Z_{N}(S_{N})$ can be expressed in terms of the ground-state energy as
  \begin{eqnarray}
   Z_{N}(S_{N}) \simeq \exp(\beta x_{N}(S_{N})) 
    = z^{x_{N}(S_{N})},
  \end{eqnarray}
  where $x_{N}(S_{N})$ is the ground-state energy with the edge spin
  state $S_{N}$.
  If we use the expressions  $x_{N}(+1)=x_{N}$ and
  $x_{N}(-1)=x_{N}+2a_{N}$, we have in the zero temperature limit
  \begin{eqnarray}
   \left[
    \begin{array}{c}
     Z_{n}^{+}\\
     Z_{n}^{-}
    \end{array}
   \right]
    \simeq
   \left[
    \begin{array}{l}
     z^{x_n}\\
     z^{x_n +2a_n}
    \end{array}
   \right].
   \label{eqn:gs}
  \end{eqnarray}
  Substituting eqn. (\ref{eqn:gs}) into eqn. (\ref{eqn:TM}), we obtain
  the following recursion relations,
  \begin{eqnarray}
   x_{n+1} &=& x_{n} +J +h_{n+1} +2\,{\rm max}(0, a_{n}-J)\label{RR:GS}\\
   a_{n+1} &=& f(a_{n}) -h_{n+1},\label{RR:ED}
  \end{eqnarray}
  where $f(a_{n})$ is a piecewise linear function defined as
  \begin{eqnarray}
   f(a_{n}) = 
    \left\{
     \begin{array}{ccl}
      +J & \textrm{for} & a_{n} > J\\
      a_{n} & \textrm{for} & -J< a_{n} < J\\
      -J & \textrm{for} & a_{n} < -J
     \end{array}
    \right. .
  \end{eqnarray}
  As shown below, from the recursion relation (\ref{RR:GS}), the
  average of the ground-state energy over random fields reduces to
  the average of ${\rm max}(0,a_{n}-J)$ over the distribution of the
  energy difference $a_{n}$.
  We note that recursion relations (\ref{RR:GS}) and (\ref{RR:ED})
  always hold for any types of random fields, not necessarily of
  eqn. (\ref{eqn:three}).

  \subsection{Ground-state energy}
  Using the recursion relation (\ref{RR:GS}), we can express the
  ground-state energy per spin as
  \begin{eqnarray}
   E_{\rm GS} &=& \lim_{N\rightarrow\infty}
    \frac{1}{N}
    \left( -\frac{1}{\beta}\log Z_{N}\right)\nonumber\\
   &=& -\lim_{N\rightarrow\infty}\frac{1}{N}
    \left\{x_{N} +{\rm max}(0,2a_{N})\right\}\nonumber\\
   &=& -\lim_{N\rightarrow\infty}\frac{1}{N}
    \left\{\sum_{n=0}^{N-1}(x_{n+1} -x_{n}) +x_{0} 
     +{\rm max}(0,2a_{N})\right\}\nonumber\\
   &=& -J -(p-r)h 
    -2\left\langle\left\langle
		   {\rm max}(0,a-J)
      \right\rangle\right\rangle,\label{eqn:EGS}
  \end{eqnarray}
  where $a$ is the energy difference in the limit $N\rightarrow\infty$
  and the double brackets $\langle\langle\cdots\rangle\rangle$ mean the
  average over the probability distribution of the energy difference
  $P(a)$.
  To obtain the last line of eqn. (\ref{eqn:EGS}), we have invoked the
  self-averaging property of the energy. 
  
  According to eqn. (\ref{eqn:EGS}), we have to derive the probability
  distribution of the energy difference $P(a)$.
  The derivation of $P(a)$ is given in \S 4.
  Before going into the detailed evaluation of $P(a)$, we explain
  another important feature of the ground state of the present model.

 \section{Spin configurations in the ground state}
 In this section, we consider the spin configuration in the ground state
 by evaluating the energy for specific configuration of random fields.
 From this argument, we derive an important property of the Ising model
 in random fields.
 
 We first consider the case $q=0$ $(h_{n}=\pm h)$.
 If the random-field strength $h$ is larger than $2J$, we can easily
 obtain the ground-state spin configuration because all spins are
 forced to be parallel to the random field.
 However, for $h<2J$, the ground state has non-trivial spin
 configurations as discussed now.

 First let us consider the following random-field configuration which
 consists of the field $+h$ at all but a single site,
 \begin{eqnarray}
  \cdots + + + + + - + + + + + \cdots,\label{fig:1}
 \end{eqnarray}
 where the symbol $+$ represents the field $+h$ and $-$ refers to
 the field $-h$.
 All spins are parallel to the random field when $h$ is larger than
 $2J$.
 However, if $h<2J$, all spins are up ($S_{n}=1$).
 To see this, we evaluate the energy difference between the two spin
 configurations,
 \begin{eqnarray}
  E_{1} = +2J -h +E_{+}\\
  E_{2} = -2J +h +E_{+}\\
  \Delta E = E_{2} -E_{1} = -4J +2h,\label{eqn:delta}
 \end{eqnarray}
 where $E_{1}$ is the energy for all spins parallel to the field,
 $E_{2}$ represents the energy of all-up spin state and $E_{+}$ means
 the energy resulting from all sites except for the location of field
 $-h$.
 According to eqn. (\ref{eqn:delta}), for $h>2J$, the ground-state
 energy is $E_{1}$, and it is $E_{2}$ otherwise.
 We note that the ground-state energy is doubly degenerate at $h=2J$.

 Next we consider the following configuration of the random field which
 consists of the field $+h$ at all sites except for two adjacent sites
 \begin{eqnarray}
  \cdots + + + + + - - + + + + + \cdots.
 \end{eqnarray}
 Similar arguments as above lead to the conclusion that all spins point
 parallel to the random field if $h>J$, and all spins are up if $h<J$.
 The ground-state energy is degenerate at $h=J$.
 
 We can apply these arguments to the general case with $k$ pieces of
 successive random fields taking the value $-h$,
 \begin{eqnarray}
  \cdots + + + + + 
   \underbrace{- - - \cdots - - -}_{k} + + + + \cdots,
 \end{eqnarray}
 for which
 \begin{eqnarray}
  E_{1} = +2J -kh +E_{+}\\
  E_{2} = -2J +kh +E_{+}\\
  \Delta E = E_{2} -E_{1} = -4J +2kh,
 \end{eqnarray}
 where the definitions of $E_{1}$, $E_{2}$ and $E_{+}$ remain the same
 as in the case of eqn. (\ref{fig:1}).
 The ground-state energy changes from $E_{1}$ to $E_{2}$ at $h=2J/k$ and
 these two energies are degenerate at $h=2J/k$.
 Consequently, the ground state of the one-dimensional Ising model in
 the random field changes at the random-field strength $h=2J/k$, where
 $k$ is an integer.
 
 To understand it, we consider the following configuration of the random
 field,
 \begin{eqnarray}
 \cdots + + + 
  \underbrace{-}_{A} + + +\cdots + + + 
  \underbrace{- -}_{B} + + + \cdots + + + 
  \underbrace{- - -}_{C} + + + \cdots,
  \label{fig:abc}
 \end{eqnarray}
 where $A$, $B$ and $C$ are the clusters of fields pointing to the
 $-$ direction with length one, two and three.
 There are several clusters of fields taking the value $-h$ in the
 sea of fields taking the value $+h$.
 In the case of $h>2J$, all spins are parallel to random fields.
 When $h$ decreases and satisfies $h<2J$, the spin at the cluster with
 length one (that is, the part $A$ in (\ref{fig:abc}))
 flips its direction from down to up.
 Next, spins at the cluster of length two (the part $B$ in
 (\ref{fig:abc})) flip their direction from down to up when $h$
 decreases and satisfies $h<J$.
 The ground-state proceeds with this change, and spins in the cluster
 with length $k$ flip their direction from $-h$ to $+h$ when the
 random-field strength reaches $h=2J/k$.
 The ground-state spin configurations are invariant in the range 
 \begin{eqnarray}
   \frac{2}{k+1}J < h < \frac{2}{k}J.\label{neq:RF}
 \end{eqnarray}
 Within this range the ground-state energy is a linear function of the
 random-field strength $h$.
 The same discussion as above holds if we include the possibility of
 random-field value of $0$.

 \section{Probability distribution}
 For explicit evaluation of the ground-state energy, we should find
 the probability distribution of the energy difference $P(a)$ according
 to eqn. (\ref{RR:ED}).
 In this section, we calculate this probability distribution $P(a)$ by
 the recursion relation (\ref{RR:ED}) following the idea of
 Refs. [\citen{DAL95}] and [\citen{KNN96}].
 
  \subsection{From recursion relation to transition matrix}
  Let us see how the $a$'s are generated stochastically by the recursion
  relation (\ref{RR:ED}).
  For example, we first consider the simplest case of the random field
  strength $h>2J$.
  We start the recursion relation from the initial condition $a_{0}=0$
  and obtain $a_{1}=-h$ for $h_{1}=h$, $a_{1}=0$ for
  $h_{1}=0$ and $a_{1}=h$ for $h_{1}=-h$.
  The next value of $a_{n}$ is 
  \begin{eqnarray}
   \begin{array}{lcll}
   a_{2} = -J-h & \textrm{for} & a_{1}=-h, & h_{2}= h\\
   a_{2} = -J   & \textrm{for} & a_{1}=-h  & h_{2}= 0\\
   a_{2} = -J+h & \textrm{for} & a_{1}=-h, & h_{2}=-h\\
   a_{2} =   -h & \textrm{for} & a_{1}= 0, & h_{2}= h\\
   a_{2} =    0 & \textrm{for} & a_{1}= 0, & h_{2}= 0\\
   a_{2} =   +h & \textrm{for} & a_{1}= 0, & h_{2}=-h\\
   a_{2} =  J-h & \textrm{for} & a_{1}= h, & h_{2}= h\\
   a_{2} =  J   & \textrm{for} & a_{1}= h, & h_{2}= 0\\
   a_{2} =  J+h & \textrm{for} & a_{1}= h, & h_{2}=-h
   \end{array}.
  \end{eqnarray}
  No other values emerge for $a_{n}$ ($n\ge 3$) as can be verified by
  repeating this procedure.
  Consequently, we can regard the recursion relation (\ref{RR:ED}) as
  a stochastic process with the transition matrix
  \begin{eqnarray}
   S_{0} =
    \left[
     \begin{array}{ccc}
      P+Q& P& P  \\
      0  & Q& 0  \\
      R  & R& Q+R\\
     \end{array}
    \right],\label{MT:S0}
  \end{eqnarray}
  where 
  \begin{eqnarray}
   P=\left[
      \begin{array}{ccc}
       p& 0& 0\\
       q& 0& 0\\
       r& 0& 0\\
      \end{array}
     \right],\hspace*{5mm}
   Q=\left[
      \begin{array}{ccc}
       0& p& 0\\
       0& q& 0\\
       0& r& 0\\
      \end{array}
     \right],\hspace*{5mm}
   R=\left[
      \begin{array}{ccc}
       0& 0& p\\
       0& 0& q\\
       0& 0& r\\
      \end{array}
     \right].
  \end{eqnarray}
  The row and column of matrix (\ref{MT:S0}) correspond to the following
  vector of the energy differences
  \begin{eqnarray}
   \left[
   \begin{array}{c}
    -J-h\\
    -J  \\
    -J+h\\
      -h\\
       0\\
      +h\\
     J-h\\
     J  \\
     J+h\\
   \end{array}
   \right].
  \end{eqnarray}
  For example, the $(1, 1)$-component of matrix (\ref{MT:S0}) means that
  the energy difference $-J-h$ remains unchanged with probability $p$.
  The stationary distribution $P(a)$ is given by the eigenvector of this
  transition matrix with the eigenvalue 1, and the result is 
  \begin{eqnarray}
   \left[
   \begin{array}{c}
    P(-J-h)\\
    P(-J)  \\
    P(-J+h)\\
      P(-h)\\
       P(0)\\
      P(+h)\\
     P(J-h)\\
     P(J)  \\
     P(J+h)\\
   \end{array}
   \right] =
   \left[
    \begin{array}{c}
     \frac{p^2}{p+r}\\
     \frac{pq}{p+r}\\
     \frac{pr}{p+r}\\
     0\\
     0\\
     0\\
     \frac{pr}{p+r}\\
     \frac{qr}{p+r}\\
     \frac{r^2}{p+r}\\
    \end{array}
   \right].
  \end{eqnarray}
  This result holds in the region $h>2J$, the case of $k=0$ in
  eqn. (\ref{neq:RF}).
  The ground state is invariant in the region of $h>2J$.
  This means that the stationary state of the range $h>2J$ is always
  described by the transition matrix $S_{0}$ and the probability
  distribution $P(a)$ remains invariant throughout the range $h>2J$.
  A similar property holds in other ranges specified by
  eqn. (\ref{neq:RF}).
  
  Applying the same argument to the case of $k\neq 0$, we obtain the 
  following $3(2k+3)\times 3(2k+3)$ transition matrix
  \begin{eqnarray}
   S_{k} =
    \left[
     \begin{array}{cccccc|c|cccccc}
      P+Q & P & 0 & \cdots & 0 & 0 & P & P & & & & &\\
      R   & Q & P & \cdots & 0 & 0 & 0 &   & & & & &\\
      0   & R & Q &        & 0 & 0 & 0  &   & & &\text{\huge{0}} & &\\
      \vdots &\vdots &   & \ddots &\vdots &\vdots & \vdots & & & & & &\\
      0  &    0  &    0  & \cdots & Q & P & 0 & & & & & &\\
      0  &    0  &    0  & \cdots & R & Q & 0 & & & & & &\\
      \hline
      0 & 0 & 0 & \cdots & 0 & 0 & Q & 0 & 0 & \cdots & 0 & 0 & 0\\
      \hline
      & & & & & & 0 & Q & P & \cdots & 0 & 0 & 0 \\
      & & & & &   & 0 & R & Q & \cdots & 0 & 0 & 0 \\
      & & & & &   & \vdots & \vdots & \vdots& \ddots & & \vdots& \vdots\\
      & & \text{\huge{0}} & & & & 0 & 0 & 0 & \cdots & Q & P & 0 \\
      & & & & & & 0 & 0 & 0 & \cdots & R & Q & P \\
      & & & & & R & R &    0  & 0 &\cdots & 0 & R & Q+R \\
     \end{array}   
    \right]. \label{MT:4k4k}
  \end{eqnarray}
  The row and column of the matrix $S_{k}$ correspond to the
  following vector of the energy differences,
  \begin{eqnarray}
   \left[
    \begin{array}{c}
     -J-h\\
     -J  \\
     -J+h\\
     -J+h-h\\
     -J+h  \\
     -J+h+h\\
     \vdots\\
     -J+kh-h\\
     -J+kh  \\
     -J+kh+h\\
     -h\\
     0\\
     +h\\
     J-kh-h\\
     J-kh  \\
     J-kh+h\\
     \vdots\\
     J-h-h\\
     J-h  \\
     J-h+h\\
     J-h\\
     J  \\
     J+h\\
    \end{array}
   \right].
  \end{eqnarray}
  Evaluation of the eigenvector of the matrix $S_{k}$ is not simple
  because this matrix includes $3\times 3$ matrices $P$, $Q$ and $R$.
  This difficulty can be resolved by the decomposition of the
  eigenvector into a direct product.
  The details are explained in Appendix.
  We only show the result here.
  The eigenvector of the matrix $S_{k}$ with eigenvalue 1 is given as
  \begin{eqnarray}
   \Vect{v} = \Vect{v_{0}}\otimes
    \left(
    \begin{array}{c}
     p\\
     q\\
     r\\
    \end{array}\right),\label{eqn:nice}
  \end{eqnarray}
  where $\Vect{v_{0}}$ is the eigenvector of the $(2k+3)\times (2k+3)$
  matrix
  \begin{eqnarray}
   S_{k} =
    \left[
     \begin{array}{cccccc|c|cccccc}
      p+q & p & 0 & \cdots & 0 & 0 & p & p & & & & &\\
      r   & q & p & \cdots & 0 & 0 & 0 &   & & & & &\\
      0   & r & q &        & 0 & 0 & 0  &   & & & \text{\huge{0}} & &\\
      \vdots &\vdots &   & \ddots &\vdots &\vdots & \vdots & & & & & &\\
      0  &    0  &    0  & \cdots & q & p & 0 & & & & & &\\
      0  &    0  &    0  & \cdots & r & q & 0 & & & & & &\\
      \hline
      0 & 0 & 0 & \cdots & 0 & 0 & q & 0 & 0 & \cdots & 0 & 0 & 0\\
      \hline
      & & & & & & 0 & q & p & \cdots & 0 & 0 & 0 \\
      & & & & & & 0 & r & q & \cdots & 0 & 0 & 0 \\
      & & & & & & \vdots & \vdots & \vdots& \ddots & & \vdots& \vdots\\
      & & \text{\huge{0}} & & & & 0 & 0 & 0 & \cdots & q & p & 0 \\
      & & & & & & 0 & 0 & 0 & \cdots & r & q & p \\
      & & & & & r & r &    0  & 0 &\cdots & 0 & r & q+r \\
     \end{array}   
    \right]. \label{MT:2k2k}
  \end{eqnarray}
  with eigenvalue 1.
    
  \subsection{Probability distribution}
  Diagonalizing the transition matrix (\ref{MT:4k4k}), we obtain the
  probability of the energy difference $P(a)$.
  For this purpose, the relation (\ref{eqn:nice}) is useful.
  We define the following eigenvector of the matrix (\ref{MT:2k2k}),
  \begin{eqnarray}
   \Vect{v_{0}}^{T} = (D_{1}, D_{2},\cdots,D_{k},D_{k+1},0,
    D_{\overline{k+1}}, D_{\overline{k}},
    \cdots,D_{\overline{2}},D_{\overline{1}}).
  \end{eqnarray}
  Using this expression of the vector $\Vect{v_{0}}$, we obtain the
  following results
  \begin{eqnarray}
   D_{m} = 
    \frac{\displaystyle \sum_{l=m}^{k+1}\left(\frac{r}{p}\right)^{l-1}}
    {\displaystyle \sum_{l=1}^{k+1}\left(\frac{r}{p}\right)^{l-1}
    \sum_{l=1}^{k+2}\left(\frac{r}{p}\right)^{l-1}}, \hspace*{5mm}
   D_{\overline{m}} = 
    \frac{\displaystyle \sum_{l=m}^{k+1}\left(\frac{p}{r}\right)^{l-1}}
    {\displaystyle \sum_{l=1}^{k+1}\left(\frac{p}{r}\right)^{l-1}
    \sum_{l=1}^{k+2}\left(\frac{p}{r}\right)^{l-1}},
  \end{eqnarray}
  where integer $m$ runs from 1 to $k+1$ and is specified by the range
  of the random-field strength as in eqn. (\ref{neq:RF}).
  The probability distribution of the energy difference $P(a)$
  corresponds to the eigenvector $\Vect{v}$ as
  \begin{eqnarray}
   \left[
    \begin{array}{c}
     P(-J-h)\\
     P(-J)  \\
     P(-J+h)\\
     P(-J+h-h)\\
     P(-J+h)  \\
     P(-J+h+h)\\
     \vdots\\
     P(-J+kh-h)\\
     P(-J+kh)  \\
     P(-J+kh+h)\\
     P(-h)\\
     P(0)\\
     P(+h)\\
     P(J-kh-h)\\
     P(J-kh)\\
     P(J-kh+h)\\
     \vdots\\
     P(J-h-h)\\
     P(J-h)\\
     P(J-h+h)\\
     P(J-h)\\
     P(J)\\
     P(J+h)\\
    \end{array}
   \right]
   =
   \left[
    \begin{array}{c}
     pD_{1}\\
     qD_{1}\\
     rD_{1}\\
     pD_{2}\\
     qD_{2}\\
     rD_{2}\\
     \vdots\\
     pD_{k+1}\\
     qD_{k+1}\\
     rD_{k+1}\\
     0\\
     0\\
     0\\
     pD_{\overline{k+1}}\\
     qD_{\overline{k+1}}\\
     rD_{\overline{k+1}}\\
     \vdots\\
     pD_{\overline{2}}\\
     qD_{\overline{2}}\\
     rD_{\overline{2}}\\
     pD_{\overline{1}}\\
     qD_{\overline{1}}\\
     rD_{\overline{1}}
    \end{array}
       \right].\label{eqn:final}
  \end{eqnarray}
  This is the final expression of the probability distribution of the
  energy difference $P(a)$.

 \section{Exact energy}  
\begin{figure}[h]
  \begin{center}
   \includegraphics{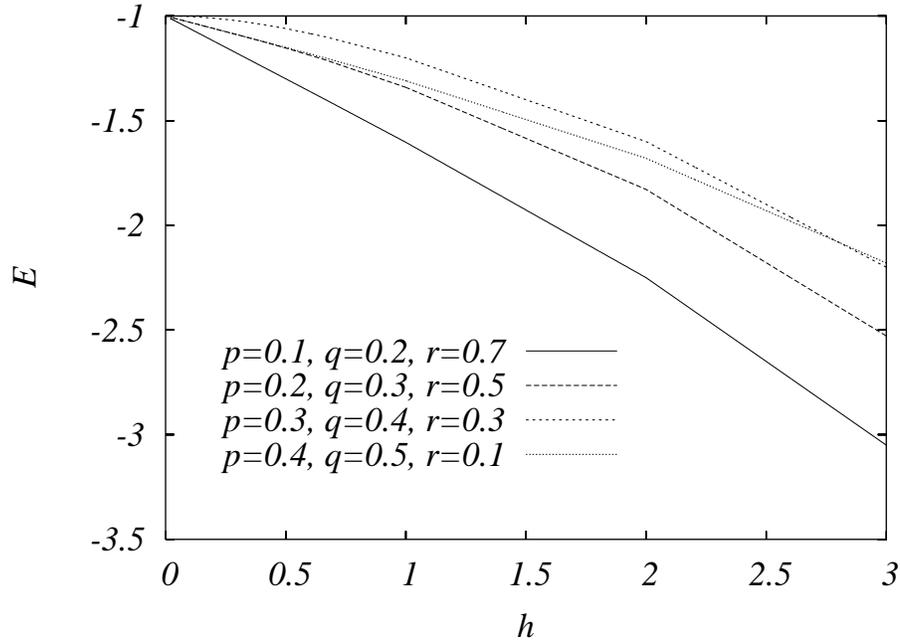}
   \caption{The ground-state energy of the random-field Ising chain with
   $h_{n}=\pm h$ and 0.}
   \label{fig:er}
  \end{center}
 \end{figure}

 Finally, we obtain the ground-state energy for $2J/(k+1)<h<2J/k$ using
 eqn. (\ref{eqn:final}) as
  \begin{eqnarray}
   E_{\rm GS}&=&
    \left\{
     -1 +\frac{4(pr)^{k+1}(p-r)^2}{A(k+1)A(k+2)}
    \right\}J\nonumber\\
   &&-
    \left\{
     1 -q +\frac{2(pr)^{k+1}(k+1)(p-r)^2}{A(k+1)A(k+2)}
     -\frac{2A(k+1)pr}{A(k+2)}
     \right\}h,\label{eqn:exg2}
  \end{eqnarray}
  where $A(k)$ is 
  \begin{eqnarray}
   A(k) = p^k -r^k.
  \end{eqnarray}
 In Fig. \ref{fig:er}, we plot the ground-state energy
 eqn. (\ref{eqn:exg2}) as a function of the random-field strength $h$ for
 several probabilities.

 \section{Ising spin ladder in bimodal random fields}
 Next, we consider the Ising spin ladder in bimodal random fields 
 $\pm h$.
 For this model, we can prove its equivalence to the one-dimensional
 model discussed in the previous sections when $h<J$.
 The exact ground-state energy of the ladder model is calculated exactly
 using the result in eqn. (\ref{eqn:exg2}).
  
 \subsection{Spin configuration for $h<J$}
 It is possible to show that for the random-field strength $h$ smaller
 than $J$, the Ising spin ladder in the bimodal random field is
 equivalent to the one-dimensional Ising model in random fields $\pm h$
 and 0.

\begin{figure}[t]
  \begin{center}
   \includegraphics[width=14cm]{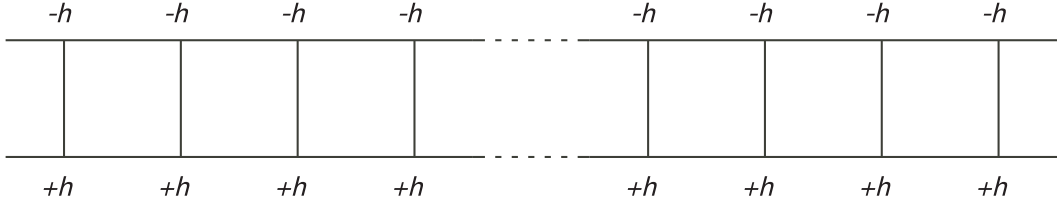}
   \caption{The configuration of fields being $-h$ in the
   first chain and $+h$ in the second chain.}\label{fig:srfc}
  \end{center}
 \end{figure}

\begin{figure}[t]
  \begin{center}
   \includegraphics[width=14cm]{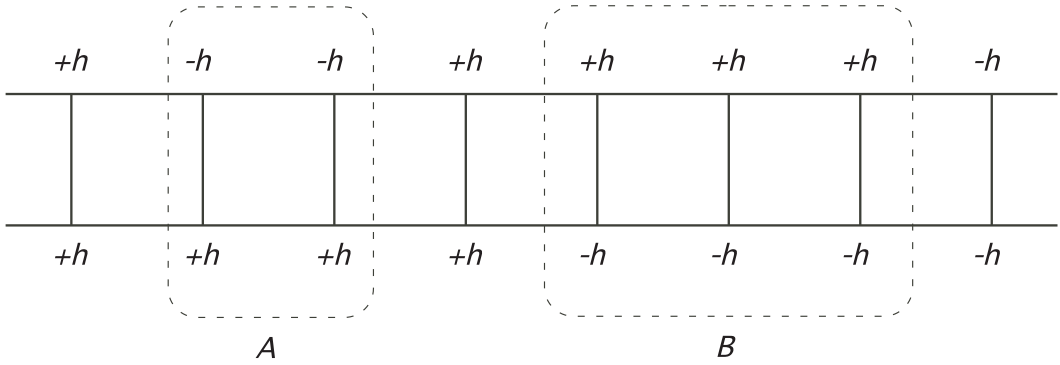}
   \caption{The general random-field configuration. The spins in parts
   $A$ and $B$ should be parallel for $h<J$.}\label{fig:grfc}
  \end{center}
 \end{figure}

 Let us consider the specific random-field configuration which consists
 of the fields being $-h$ in the first chain  and $+h$ in the second
 chain (Fig. \ref{fig:srfc}).
 For this random-field configuration, there are two possible
 ground states.
 One consists of all spins being up (or down due to the symmetry), and
 the other is all spins being parallel to the field.
 We introduce the notation $E_{\rm allup}$ for the energy of the first
 ground state and $E_{\rm fields}$ for the energy of the second ground
 state.
 The energies $E_{\rm allup}$ and $E_{\rm fields}$ are 
 \begin{eqnarray}
  E_{\rm allup} &=& -2(n-1)J -nh +n(h-J)\\
  E_{\rm fields} &=& -2(n-1)J -nh +n(J-h),
 \end{eqnarray}
 where $n$ is the number of the rungs.
 If the random-field strength $h$ is smaller than $J$, the ground-state
 energy becomes $E_{\rm allup}$, and the ground state is all spins being
 up or down with equal probability.
 Let us consider a general random-field configuration as shown in
 Fig. \ref{fig:grfc}.
 All spins in the parts $A$ and $B$ should be up or down from the above
 discussion.
 The spins in the left-most rung should be up because the fields on this
 rung are $+h$, and the spins between the parts $A$ and $B$ are also up
 for the same reason as the left-most rung.
 Accordingly, the spins in the part $A$ should be up.
 The spins in the right-most rung should be down for the fields on this
 rung being down, and there are two degenerate states between all spins
 in the part $B$ being up and down.
 The same argument as above holds for any other random-field
 configurations.
 For any case, the spins in the same rung are parallel.
 Consequently, we conclude that for $h<J$, two spins on the same rung in
 the ladder model always point to the same direction.
 Therefore, by identifying spins on the same rung as a spin taking the
 values $\pm 1$, we obtain the effective coupling constant $2J$ and the
 three kinds of effective fields $\{+2h, 0, -2h\}$.
 This means that the ladder model in the bimodal random field is
 equivalent to the Ising chain in random fields $\pm h$ and 0.

 \subsection{Exact energy}
 From above argument, the exact ground-state energy of the ladder model
 can be obtained from the result in eqn. (\ref{eqn:exg2}).
 The corresponding probabilities of the random field are 
 \begin{eqnarray}
  \begin{array}{lcc}
   p = x^2 & {\rm for} & +h\\
   q = 2x(1-x) & {\rm for} & 0\\
   r = (1-x)^2 & {\rm for} & -h\\
 \end{array},
 \end{eqnarray}
 where $x$ means the probability of the random field being $+h$ for the
 ladder model.
 We also replace the random field strength and the coupling constant
 of the one-dimensional model with $J \rightarrow 2J$ and 
 $h \rightarrow 2h$.
 We need an additional term $J$ which is caused by spins on the same
 rung pointing to the same direction, and divide the result by 2 because
 the number of spins in ladder model is twice as many as the
 one-dimensional model.
 The final expression is
  \begin{eqnarray}
   E_{\rm GS}&=&
    \left\{
     -\frac{3}{2} +\frac{4(pr)^{k+1}(p-r)^2}{A(k+1)A(k+2)}
    \right\}J\nonumber\\
   &&-
    \left\{
     1 -q +\frac{2(pr)^{k+1}(k+1)(p-r)^2}{A(k+1)A(k+2)}
     -\frac{2A(k+1)pr}{A(k+2)}
     \right\}h,\label{eqn:exg2leg}
  \end{eqnarray}
  where $A(k)$ is 
  \begin{eqnarray}
   A(k) = p^k -r^k.
  \end{eqnarray}

 \section{Discussion}
 In the present paper, we have derived the exact ground-state energy of
 the random-field Ising chain with field values $\pm h$ and 0 using the
 zero-temperature transfer matrix method in
 Refs. [\citen{DAL95}] and [\citen{KNN96}].
 We have also obtained the exact ground-state energy of the Ising ladder
 in bimodal random fields for $h<J$.
 As far as we know this is the first example in which an exact solution
 has been derived for ladder model in random fields.
 The present techniques is a step toward a solution of random-field
 Ising models on strips with wider widths, eventually reaching the
 two-dimensional model.

 We give a few comments on some aspects other than the ground-state
 energy.
 First, it should be possible to derive the exact solution for other
 physical quantities than the ground-state energy, such as the entropy
 and magnetization, using the methods in Refs. [\citen{DVP78}] and
 [\citen{R93}].
 Investigations in this direction are under way.

 For the Ising ladder in bimodal random fields, the evaluation of the
 transition matrix is not straightforward for $h>J$ since we cannot map
 the problem to the one-dimensional model.
 This difficulty is caused by the complicated structure of the recursion
 relation of the ladder model.
 In fact, we find four kinds of recursion relations to evaluate the
 probability distribution of the energy differences.
 It is not easy to construct the general form of the transition matrix
 from these recursion relations. 
 This is nevertheless an important and challenging task because of the
 eventual target of the two-dimensional model as mentioned above.

 \section*{Acknowledgments}
 We would like to thank A. C. C. Coolen and P. Contucci for useful
 comments.
 We would also like to thank T. Sasamoto for useful comments to
 derive the argument in Appendix and Y. Ozeki, K. Takeda and S. Morita
 for interesting discussions.
 This work was supported by the 21st Century COE Program at Tokyo
 Institute of Technology ``Nanometer-Scale Quantum Physics'' and the
 Grand-in-Aid for Scientific Research on Priority Area
 ``Statistical-Mechanical Approach to Probabilistic Information
 Processing'' by the Ministry Education, Culture, Sports, Science and
 Technology.

\appendix
 \section{Decomposition of eigenvectors into a direct product}
 The matrix (\ref{MT:4k4k}) is not simple to diagonalize because it is
 composed of the $3\times 3$ matrices $P$, $Q$ and $R$.
 To diagonalize this $3(2k+3)\times 2(2k+3)$ matrix, we use the
 decomposition of the eigenvector into a direct product.
 We define the following vector $\Vect{w}$,
 \begin{eqnarray}
  \Vect{w} = \left(
	     \begin{array}{c}
	      p\\
	      q\\
	      r\\
	     \end{array}
	    \right).
 \end{eqnarray}
 This vector satisfies the following relations
 \begin{eqnarray}
  P\Vect{w} = p\Vect{w}, \hspace*{5mm} Q\Vect{w} = q\Vect{w}.\label{eqn:ePQ}
 \end{eqnarray}
 Using eqn. (\ref{eqn:ePQ}), we can decompose the eigenvector $\Vect{v}$
 of the matrix $S_{k}$ into a direct product of vectors $\Vect{v_0}$
 and $\Vect{w}$ as
 \begin{eqnarray}
  \Vect{v} = \Vect{v}_{0}\otimes\Vect{w},\label{eqn:vect}
 \end{eqnarray}
 where $\Vect{v_0}$ is the eigenvector of matrix $\tilde{S}_{k}$ in
 eqn. (\ref{MT:2k2k}) with eigenvalue 1.
  To prove eqn. (\ref{eqn:vect}), we show that the vector
 $\Vect{v_0}\otimes\Vect{w}$ satisfies the equation for the eigenvector
 of the matrix $S_{k}$ with eigenvalue 1 as
 \begin{eqnarray}
  S_{k}\Vect{v} = S_{k}(\Vect{v}_{0}\otimes\Vect{w}) 
   = (\tilde{S}_{k}\Vect{v}_{0})\otimes\Vect{w}   
   = \Vect{v}_{0}\otimes\Vect{w}
   = \Vect{v}.
 \end{eqnarray}
 Thus eqn. (\ref{eqn:vect}) holds.
 Accordingly, it is useful for the evaluation of the probability
 distribution that we calculate the eigenvector of the matrix
 $\tilde{S}_{k}$ with eigenvalue 1 instead of the eigenvector of the
 matrix $S_{k}$ with eigenvalue 1.


\begin{thebibliography}{99}
   \bibitem{DVP78} B. Derrida, J. Vannimenus and Y. Pomeau:
	   J. Phys. C {\bf 11} (1978) 4749.
   \bibitem{FG93} E. Farhi and S. Gutmann:
	   Phys. Rev. B {\bf 48} (1993) 9508.
   \bibitem{V78} A. Vilenkin:
	   Phys. Rev. B {\bf 18} (1978) 1474.
   \bibitem{DAL95} C. Dress, E. Amic and J. M. Luck:
	   J. Phys. A: Math. Gen. {\bf 28} (1995) 135.
   \bibitem{KNN96} T. Kadowaki,Y. Nonomura and H. Nishimori:
	   J. Phys. Soc. Jpn. {\bf 65}  (1996) 1609.
   \bibitem{S96} P. Shukla:
	   Physica A {\bf 233} (1996) 235.
   \bibitem{S00} P. Shukla:
	   Phys. Rev. E {\bf 62} (2000) 4725.
   \bibitem{DDMZ02} L. Dante, G. Durin, A. Magni and S. Zapperi: 
	   Phys. Rev. B {\bf 65} (2002) 144441.
   \bibitem{R93} P. Rujan:
	   Phys. Rev. Lett {\bf 70} (1993) 2968.
  \end{thebibliography}
\end{document}